\documentclass[10pt,journal,compsoc]{IEEEtran}

\usepackage{microtype}                 
\PassOptionsToPackage{warn}{textcomp}  
\usepackage{textcomp}                  
\usepackage{mathptmx}                  
\usepackage{times}                     
\usepackage{cite}                      
\usepackage{tabu}                      
\usepackage{booktabs}                  
\usepackage{amsmath}
\usepackage{amssymb}
\usepackage{amsfonts}
\usepackage{multirow}
\usepackage[table,xcdraw]{xcolor}
\usepackage{enumitem}
\usepackage{graphicx}
\usepackage[colorlinks,bookmarksopen,bookmarksnumbered,citecolor=red,urlcolor=red]{hyperref}
\usepackage[switch]{lineno}
\usepackage{subfig}

\graphicspath{{figures/}{pictures/}{images/}{./}}

\begin{document}


\title{CoronaViz: Visualizing Multilayer Spatiotemporal COVID-19 Data with Animated Geocircles}

\author{Brian Ondov, Harsh B. Patel, Ai-Te Kuo, Hanan Samet,~\IEEEmembership{Life Fellow, IEEE}, John Kastner,\\ Yunheng Han, Hong Wei, and Niklas Elmqvist,~\IEEEmembership{Senior Member, IEEE}
\IEEEcompsocitemizethanks{
    \IEEEcompsocthanksitem Brian Ondov is with the National Library of Medicine in Bethesda, MD, USA. E-mail: brian.ondov@nih.gov.
    \IEEEcompsocthanksitem Harsh B. Patel, Hanan Samet, Yunheng Han, Hong Wei, and Niklas Elmqvist are with University of Maryland in College Park, MD, USA. E-mail: \{hpatel01, hjs, yhhan, hyw, elm\}@umd.edu.
    \IEEEcompsocthanksitem Ai-Te Kuo is with Auburn University in Auburn, AL, USA. E-mail: aitekuo@auburn.edu.
    \IEEEcompsocthanksitem John Kastner is with Amazon (contribution made prior to employment). E-mail:  jkastner@amazon.com.
    \IEEEcompsocthanksitem Hong Wei is with Facebook. E-mail:  hyw@umd.edu.}
    \thanks{Manuscript received XXX XX, 2022; revised XXX XX, 2022.}
}

\IEEEtitleabstractindextext{%
\begin{abstract}
While many dashboards for visualizing COVID-19 data exist, most separate geospatial and temporal data into discrete visualizations or tables. Further, the common use of choropleth maps or space-filling map overlays supports only a single geospatial variable at once, making it difficult to compare the temporal and geospatial trends of multiple, potentially interacting variables, such as active cases, deaths, and vaccinations. We present CoronaViz, a COVID-19 visualization system that conveys multilayer, spatiotemporal data in a single, interactive display. CoronaViz encodes variables with concentric, hollow circles, termed geocircles, allowing multiple variables via color encoding and avoiding occlusion problems. The radii of geocircles relate to the values of the variables they represent via the psychophysically determined Flannery formula. The time dimension of spatiotemporal variables is encoded with sequential rendering. Animation controls allow the user to seek through time manually or to view the pandemic unfolding in accelerated time. An adjustable time window allows aggregation at any granularity, from single days to cumulative values for the entire available range. In addition to describing the CoronaViz system, we report findings from a user study comparing CoronaViz with multi-view dashboards from the New York Times and Johns Hopkins University. While participants preferred using the latter two dashboards to perform queries with only a geospatial component or only a temporal component, participants uniformly preferred CoronaViz for queries with both spatial and temporal components, highlighting the utility of a unified spatiotemporal encoding. CoronaViz is open-source and freely available at  \url{http://coronaviz.umiacs.io}.
\end{abstract}

}

\maketitle
\IEEEdisplaynontitleabstractindextext
\IEEEpeerreviewmaketitle

\begin{figure*}[htb]
    \centering
    \includegraphics[width=\linewidth]{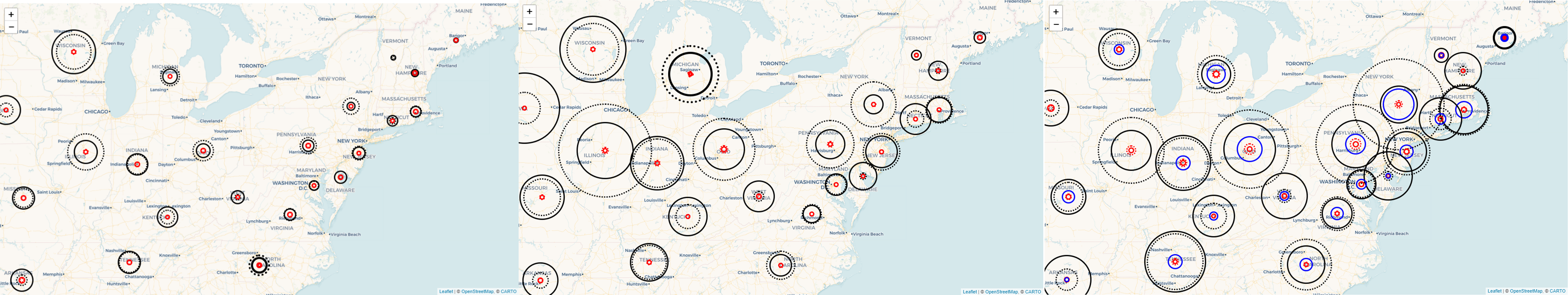}
    \caption{\textbf{Spatiotemporal visualization using geocircles.}
    Shown are selected frames from CoronaViz's animated spatiotemporal visualization, which encodes spatial variables with location-centered, hollow, concentric circles. Time is progressing from left to right.}
    \label{fig:teaser}
\end{figure*}

\section{Introduction}

\IEEEPARstart{D}{ata} visualization is an integral part of the modern study of epidemiology~\cite{carroll2014visualization}.
This role dates at least to 1854, when John Snow famously plotted cholera cases on a map of London, helping to end an outbreak as well as advance the germ theory of disease~\cite{mcleod2000our}.\footnote{Though the common causal narrative has been questioned, cartographic visualizations likely played a crucial role in this epidemic nonetheless.}
In particular, the concepts of space and time are central to the detective work associated with contact tracing during a disease outbreak~\cite{wang2020public}.
Today, epidemiologists have a wealth of digital tools at their disposal, thanks both to worldwide data collection and to advances in Geographic Information Systems (GIS) and other interactive forms of data visualization.
Accordingly, visualization has played a crucial role in understanding and combating the novel coronavirus disease COVID-19 since it was first declared a pandemic~\cite{boulos2020geographical, bowe2020learning, comba2020data, leung2020big}.

There exist numerous visualization and GIS systems to monitor officially released numbers of cases~\cite{dong2020interactive}, which are the current established means of keeping track of the progress of the virus.
However, these systems do not necessarily paint a complete picture.
In particular, most use static representations aggregating data at specific time intervals or specific geographic locations.
Temporal data, if available, is often shown in a separate view, taking a dashboard approach.
Many systems also suffer from overplotting and high visual clutter, and can only represent a single geospatial variable at a time.

We present \textsc{CoronaViz} (\href{https://coronaviz.umiacs.io}{http://coronaviz.umiacs.io}), a spatiotemporal visualization using a dynamic map of COVID-19 data including the number of confirmed cases, active cases, recoveries, deaths, and vaccinations.
The data is drawn from the Johns Hopkins University website.
CoronaViz overcomes many of the issues with other tools by combining existing visualization techniques, namely, animation of temporal data and the use of hollow proportional symbols, in this case concentric circles, which we term \textit{geocircles}.
The result is a unified spatiotemporal encoding of multiple variables, enabling the user to observe the interactions of variables like cases, deaths, and vaccinations as they unfold on an accelerated timeline (Fig.~\ref{fig:teaser}).

This paper is a significantly extended version of work presented at the ACM SIGSPATIAL 2020 Workshop on Modeling and Understanding the Spread of COVID-19~\cite{samet2020using}.
In particular, an extensive user study was conducted, as well as an expansion of the underlying data to include vaccinations, and the improvement of the user interface based on user feedback.
We claim the following main contributions in this paper:

\begin{enumerate}
    \item An open-source, freely hosted Web application for visualizing and exploring COVID-19 data with multiple layers and temporal animation;
    
    \item The formulation and use of animated concentric \textit{geocircles} to represent spatiotemporal variation in COVID-19 related variables while avoiding overplotting; and
    
    \item Results from a user study assessing the value of animated geocircles for answering epidemiological queries with both spatial and temporal components compared to other COVID-19 tools.
\end{enumerate}


\section{Background}
\label{rw}

There are several areas that relate to displaying time-varying, quantitative information.
Here we will review quantitative geospatial visualization and the added complexity of spatiotemporal data. We will also discuss existing systems specifically for the application of tracking the COVID-19 pandemic.

\subsection{Quantitative Geospatial Visualization}

Maps that convey quantitative variables associated with regions are often called thematic maps.
Common thematic map variants include choropleth maps~\cite{dixon1972methods}, which vary hue or shading of regions according to their values, and proportional symbol maps~\cite{slocum_thematic_2009}, in which size-varied symbols are overlaid onto locations they represent. 
For dense visualizations of quantities with wide variations in magnitudes, overlaid representations often run into the problem of occlusion, in which the overlaid symbols or charts block the view of other symbols or of significant geographical features of the regions they represent.

Several methods have been proposed to overcome occlusion, including the use of ``hollow'' proportional symbols, alpha channel blending~\cite{chen2014visual}, and necklace maps~\cite{speckmann2010necklace}, which project proportional symbols onto curves surrounding a larger region.
Cartograms attach quantities to regions by distorting their borders such that their areas are proportional to a variable of interest~\cite{tobler1963geographic}.
Many variants of cartograms exist; see the survey by Nusrat and Kobourov~\cite{DBLP:journals/cgf/NusratK16}

\subsection{Spatiotemporal Data Visualization}

Visualizing spatiotemporal data on a map is a challenging prospect; as such, it has been the subject of substantial prior work.
The difficulty comes from the inherently multidimensional nature of the data: there are at a minimum two spatial dimensions and one temporal dimension, in addition to the dimensionality added by the actual variables being visualized.
All of these dimensions must be projected onto a two-dimensional screen.
Unlike traditional time-series data such as a linechart, where time can be assigned a specific Cartesian axis, a geographic map already consumes both available 2D geometric axes.
This means that time-varying data on a map generally is often visualized over time as animation. 
For a more complete survey and typology of spatiotemporal data visualization see Andrienko et al.~\cite{andrienko2003exploratory}; however, we will discuss several strategies here.

We can broadly break spatiotemporal visualization techniques into two groups: those that use animation to capture the time dimension and those that attempt to encode temporally varying information into a single static visualization.
Animation is often employed by meteorological visualizations~\cite{papathomas1988applications, schiavone1990visualizing} or for satellite observations of geological features, such as temperature or surface reflectance~\cite{bladin2017globe}.
All these systems vary the data overlaid on the map over time.
Going beyond overlays, Ouyang and Revesz develop an algorithm to generate spatiotemporal cartogram animations, shifting the area of regions according to a time-varying variable~\cite{ouyang_algorithms_2000}.

An example of the second group, temporal encoding, is presented by Du et al.~\cite{du_banded_2018}, who modify choropleth maps to encode temporal information inside each area unit.
Rather than picking a single color for each areal unit, units are divided either into bands of either equal width or equal area.
Each band is then assigned a color in the same way areal units are assigned colors in traditional choropleth maps (e.g.~\cite{slocum_thematic_2009}).
Sun et al.~\cite{sun2014embedding} embed time series charts of traffic data within street maps.
Deng et al.~\cite{deng2021compass} use ``compass'' plots alongside a map to convey temporal causality. Maciejewski et al.~\cite{maciejewski2009visual} employ ``ghosting,'' in which data from more recent times is displayed at higher opacity.
Li et al.~\cite{li_cope_2019} do not use a fully animated approach, but neither do they commit to showing the full temporal data range in a single image.
Instead, they use an interface termed the ``Event View'' to display images generated for discrete time intervals side-by-side. 
To link these images together into a single cohesive visualization, they overlay a ``trend line'' that
connects the time intervals.
This trend line is used to link events extracted by a separate component of their system.

An alternate temporal encoding is to use the third dimension for this purpose.
Space-time cubes~\cite{gatalsky2004interactive} use a three-dimensional perspective view to display information at varying heights above a section of a map, with the z-axis representing time.
Space-time cubes have also been applied specifically to COVID-19 data~\cite{mo2020analysis}.
Similarly, GeoTimes~\cite{DBLP:journals/ivs/EcclesKHW08} uses the third (upwards-pointing) vector to visualize the temporal events in 3D.

\subsection{Existing COVID-19 Monitoring Systems}

The wide availability of both data and out-of-the-box GIS solutions has yielded an explosion of dashboards specifically for geospatial visualization of COVID-19 data.
An exhaustive list of such dashboards is thus beyond the scope of this paper; see Boulos et al.~\cite{boulos2020geographical} for additional dashboards and references to further lists of dashboards. 
However, we will review here a number of existing systems that are notable or especially relevant to our work.

The New York Times' dashboard~\cite{tool-nyt} contains cumulative data displayed in tabular format, by the use of hotspot maps, and time series line charts for variables of the disease. The time series graphs show a fixed 7-day average of a particular variable. The data displayed in tabular format allows users to search for some location and sort by a particular variable either for all time or for recent trends.

The Johns Hopkins system~\cite{tool-jhu, dong2020interactive}, using the ArcGIS platform~\cite{scott2010spatial}, displays current cumulative numbers of confirmed cases, active cases, deaths, and recoveries.
The cumulative numbers of confirmed and active cases in some of the countries are displayed on the map for some of the larger countries (in terms of area). The system also displays tabular data and time series data for a focused location, forming a dashboard.

The HealthMap system~\cite{tool-healthmap} shows the spread of the disease by tabulating the number of new
confirmed cases of the disease on a daily basis and displaying it with a circle of a particular size and color anchored at the location where it was reported (e.g.,~a city, state, country, etc.). HealthMap is notable in that, like CoronaViz, it uses an integrated spatiotemporal visualization using animation. However, its animation controls are rudimentary, lacking the abilities to adjust time window, adjust playback speed, and even to pause and resume. Further, HealthMap only shows a single spatiotemporal variable (confirmed cases), and at a single granularity, which leads to severe overplotting when zoomed out.

The Google News system~\cite{tool-google} makes use of a map query interface and allows zooming in and reports the variable values for the smaller subunits.
It uses a hover operation to yield the variable values for the spatial unit being hovered over, as well as disease-related news at times. 
It does not have the ability to provide variable values for a combination of units that make up the viewing window when these units are small (e.g., counties) or bigger (countries) as is done in CoronaViz.
It is static as it has no temporal component other than precomputed graphs of variable values over a predetermined range of days unlike CoronaViz where the range is set by the user.



Both the 1point3acres~\cite{tool-1point3acres} and Worldometer~\cite{tool-worldometer} systems provide comprehensive data and graphs for the dynamic variables but no animation or maps.
The dynamic aspect of the variables is captured by the various plots of the variable values and their combinations.
They make a distinction between cumulative variable values as well as new values.
The 1point3acres system emphasizes data collection ability and is more focused on the virus while Worldometer also provides statistics related to the impact of the disease such as unemployment.


Other systems tend to be quite similar in that they only map the number of confirmed cases in each country, in the case of the WHO~\cite{tool-who} and ECDC~\cite{tool-ecdc} systems, and in each state for the CDC system~\cite{tool-cdc}.
The Kaiser Family Foundation system~\cite{tool-kaiser} also maps the deaths.
None of the WHO, ECDC, CDC, and the Kaiser Family Foundation systems permit zooming in to get additional data.
Non-interactive maps are used to tell the story of the coronavirus outbreak in the South China Post~\cite{tool-southchina} using ESRI StoryMaps~\cite{tool-esri}.
Instead of the disease-related variables, some systems such as that from the University of Wisconsin monitor the mobility of the population with a map query interface based on cellphone data.

\section{Geocircles for Animated Spatiotemporal Queries}
\label{qry}

We propose a proportional symbol map in which the symbols are hollow circles, multiple variables are encoded concentrically using color and line style variation, and in which animation is used to convey time-varying data. We refer to these symbols as \textit{geocircles}.

\subsection{Challenges: 2D Spatiotemporal Visualization}

Presenting several static variables on a map often leads to visual clutter regardless of whether they are represented as one graph for the set of all variables or one graph per variable.
The situation becomes more complex when values of the variables vary in a spatially-varying manner.
In this case the only way to deal with the static variables is to repeat the graph at each location.
This is only possible when the data is spatially sparse.

However, we need a more scalable approach when the data is not spatially sparse.
For example, a histogram representing time-varying changes leads to visual clutter and is impractical for spatially dense data.
Moreover, there may be a layout problem here in the sense that we cannot allow the histograms to overlap.
An alternative common solution with the same overlap issues is to use solid concentric circles as proportional symbols
The problem is that when there are multiple dynamic variables, as is the case for our application, then only the one with the largest magnitude can be viewed.
One solution is to vary the colors of the circles, but then the order in which the circles are rendered becomes important. This kind of ``z-indexing'' may be feasible for a static view, but the ordering may need to change in a dynamic visualization, which could leading to jarring visual artifacts.

\subsection{Geocircle Design}

Each circle is centered on its respective spatial position on the map.
The radius of the circle corresponds to a scaled variable magnitude and its color indicates the identity of the variable.
The hollow (unfilled) nature of the circles means that we don't have to worry about the order in which we display the circle.
We use the term {\em geocircle} to describe this approach. Geocircles bear resemblance to the Halo technique~\cite{DBLP:conf/chi/BaudischR03} by Baudisch and Rosenholtz, which uses hollow circles centered on off-screen map objects to make the user aware of such objects and their proximity to the screen edge.
In a sense, our intention is the same, but we use the concentric circles for on-screen targets instead. 

The visual strain posed by having a large number of circles can be relieved by drawing the circles using broken lines of the same width.
At times, the width of the broken lines can be increased with the goal of drawing attention to a particular set of concentric circles (i.e., a location whose variable values at a particular instance of time) which is of interest.
We do this in the case of a hover operation while panning on the map to show the spatially closest location with nonzero variable values.
This operation is common in computer graphics where it is known as a ``pick'' operation (e.g.,
see~\cite{Fole90}).
However, care must be exercised when implementing it in the sense that we don't always want the closest geocircle.
For example, if we are hovering over Brazil, then we want the geocircle of Brazil even though the geocircles of Paraguay or Bolivia may be closer to the hover location in Brazil.

\subsection{Spatiotemporal Data Analysis using Geocircles}

The concentric circles make it easy to spot trends and similar values on the map by looking at the magnitude of the radii.
Other observations of interest involve trends changing over time as the circles get larger or smaller.
Another encouraging trend is when confirmed case counts become smaller than death counts.
Of particular interest is the situation when concentric circles intersect and change their relative order (of course this must be treated with caution as the magnitudes of the variables change).
In particular, of a comparison is only meaningful when comparing variable values and not rates.

There are a number of ways of presenting the variable values.
The default in our case is of a cumulative nature.
However, it is possible to normalize the values over population, or even area.
Normalizing over the area is of possible interest as it could be used to see if densely populated areas are more likely to lead to a higher variable numbers. 

There are two main types of queries:
\begin{enumerate}
\item\textbf{Location-based:} given a location or time, what are the values of certain variables and rates.
\item\textbf{Feature-based:} given a variable or rate value, where or when is its value present.
This is also known as spatial data mining~\cite{Aref90a}.
\end{enumerate}

The location-based queries are supported by the ability to pan the map with a hover operation and always returning the variable values with the nearest location for which we have data.
This might be implemented, for example, through a PR quadtree for each of the dynamic queries or variables or rates.
Feature-based queries require the use of a pyramid data structure on each of the dynamic queries or variables or rates.

The animation window enables the execution of a range query where the range is temporal and spatial.
Users can vary the start and end times of the query as well as the animation step size. 
In addition, users can specify what statistic is being computed for the temporal window.
It can be cumulative, or a time period whose length can be in terms of days, weeks, months, or even years.  
Average values for the window can also be computed.
This may be particularly useful for informing evidence-based policy, as decisions to close or open businesses in particular jurisdictions are often tied to daily or weekly averages of case numbers~\cite{zhang2020covid}.


Geocircles can be used for the full compliment of spatiotemporal queries as it supports keeping location fixed while varying time via the time slider, keeping time fixed and letting location vary via the hover, panning, and zooming operations.
We can also pick any range of time or space.
Note that for temporal proximity, we provide the capability to halt an animation at arbitrary time instances as well as resuming or terminating it.
In addition, users are also able to set the speed of the animation, as well as to step through an animation by a specific time interval both forward and backward in time.

\section{CoronaViz: System Design}
\label{ui}

\begin{figure*}[ht!]
    \centering
    \includegraphics[width=\textwidth]{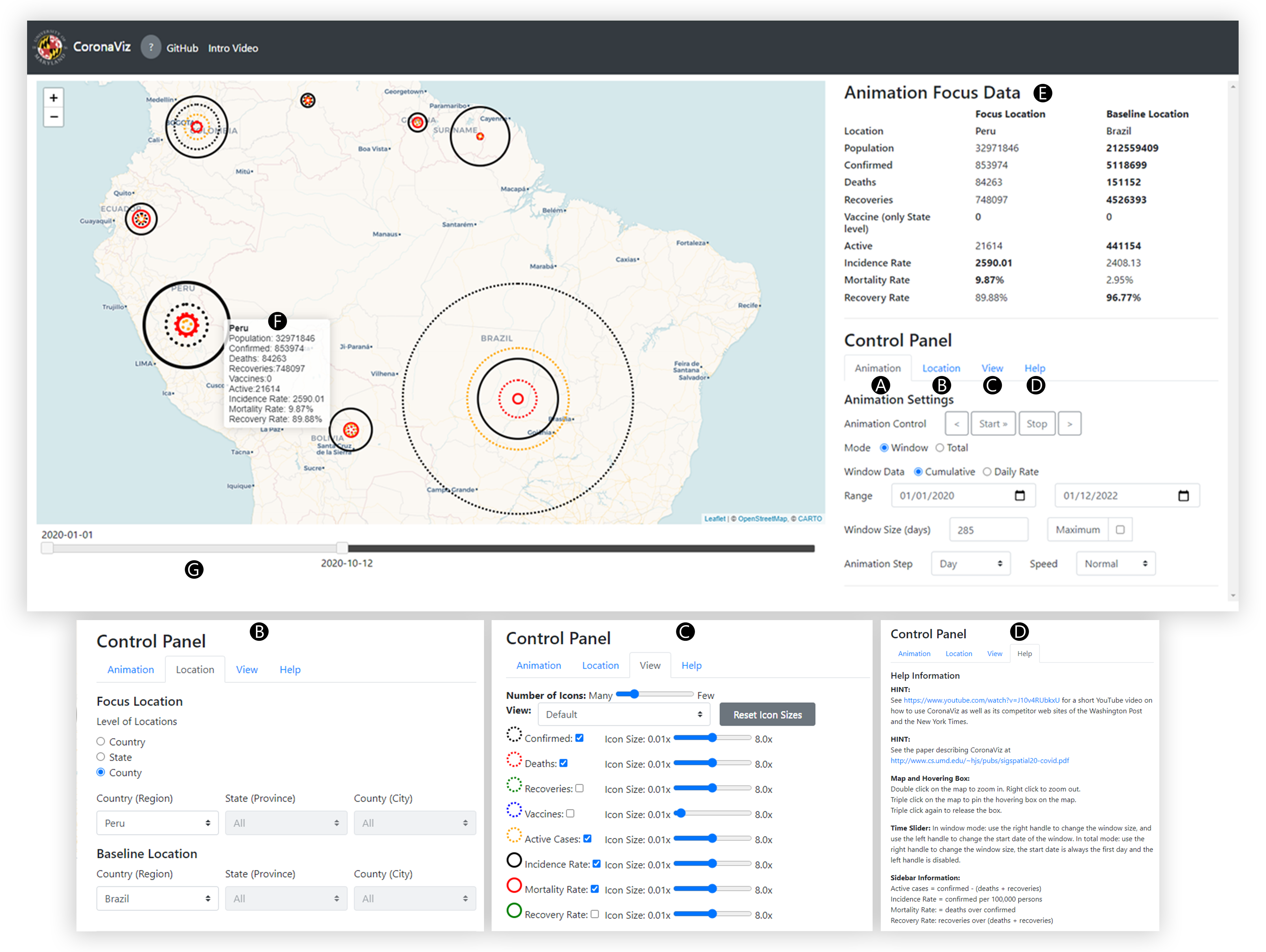}
    \caption{Overview of the CoronaViz tool and its user interface, showing COVID-19 data for Brazil and neighboring countries. The Control Panel contains four tabs: (A) Animation Settings, (B) Focus Location, (C) View settings, and (D) Help. Below, tabs (B)-(D) as they would appear in the Control Panel when selected. (E) The Animation Focus Data panel displays data for two given locations which are adjusted by clicking on the map or selecting a location in (B). (F) A hover box containing detailed information is seen when the user hovers over a region. (G) The time slider controls the start (left box) and the end (right box) dates of the windowed data.}
    \label{fig:brazil-ui-overview}
\end{figure*}

We present \textsc{CoronaViz}, a spatiotemporal visualization tool implemented as a web-based application that uses animated geocircles to visualize continuously-updating data of the COVID-19 pandemic.
Here we describe the data model, user interface, and utility.

\subsection{Data Model and Visual Encoding}

CoronaViz makes use of \textit{dynamic variables} such as confirmed cases, active cases, recoveries, deaths, and vaccinations, as well as \textit{normalized rates} that include the incidence rate (number of confirmed cases per 100,000 inhabitants), mortality rate (number of deaths divided by the number of confirmed cases), and the recovery rate (number of recoveries divided by the sum of the numbers of deaths and recoveries).
No active rate is tabulated as the number of active cases is simply the number of confirmed cases minus
the number of deaths and recoveries; thus the only possible rate measure is a normalized active cases value per 100,000 inhabitants. 
This is similar to the incidence rate and thus we do not provide it.

Concentric circles (i.e., geocircles) drawn with broken lines are used for absolute variable values while circles drawn with solid lines are used for rates.
They are drawn with different colors, with the same color being used for the corresponding variable and rate.

\subsection{Interface Overview}

Figure~\ref{fig:brazil-ui-overview} provides an overview of the CoronaViz interface. The page is divided into three main sections: the map view, the Focus Data Panel, and the Control Panel, which has four components: data animation, location specification, data viewing, and help. They are accessed by corresponding tabs.

In this case, the map is showing confirmed cases (broken black circles), deaths (broken red circles), active cases (broken yellow circles), incidence rate (solid black circles), and mortality rate (solid red circles) for a region of South America. From the time sliders (Fig. 2 G), we can see that these values are for a 285-day period spanning from January 1, 2020 through October 12 of the same year.
The result of the query is a static, cumulative view, in which geocircles represent linearly scaled incidence rate and linearly scaled mortality rate corresponding to the  countries in South America.
From the figure we see that the incidence rates are relatively similar for these countries.
Mortality rates are much smaller and thus we may wish to scale  them so that we can better differentiate between them.
Figure~\ref{fig:brazil-ui-overview} A-D show the different control panels the user can select from.
The different control panels are described in detail below.

\begin{figure}[htb!]
    \centering
    \subfloat[DC and Maryland separate]{
        \includegraphics[width=0.45\textwidth]{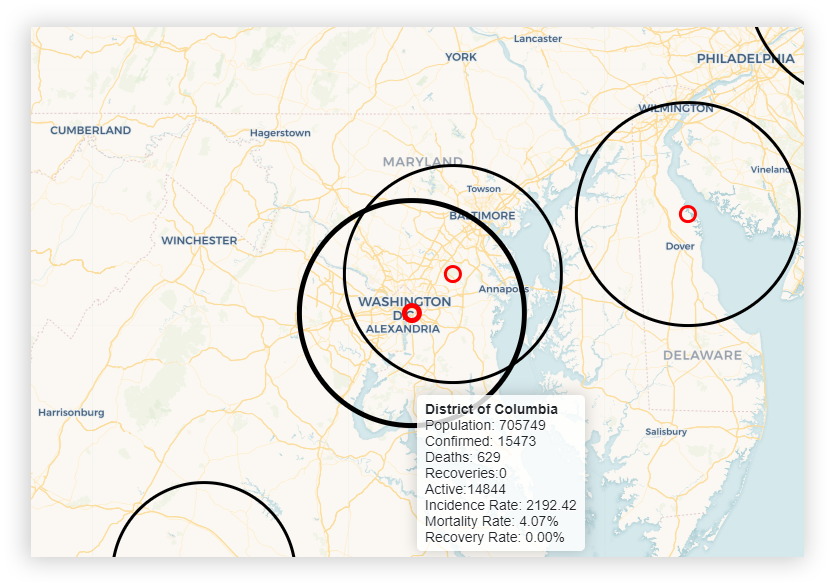}
        \label{fig:dc-maryland-separate}
    }\\
    \subfloat[DC and Maryland aggregate]{
        \includegraphics[width=0.45\textwidth]{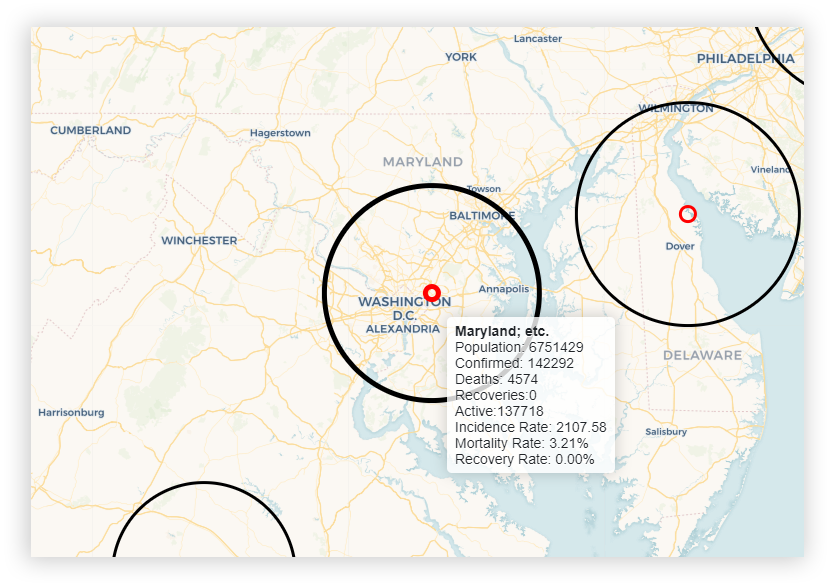}
        \label{fig:dc-maryland-aggregate}
    }
    \caption{CoronaViz reduces clutter by aggregating geocircles of nearby regions, e.g. Maryland and District of Columbia in (a), into groups, such as ``Maryland; etc.'' in (b). ~(\href{https://www.youtube.com/watch?v=DYHk5XmGXKA}{\color{blue}\underline{video}}). The amount of aggregation can be controlled by the user via the View tab of the Control Panel.}
    \label{fig:limit-number}
\end{figure}

\subsection{Animation and Spatiotemporal Queries}

The ``Animation'' tab (Fig.~\ref{fig:brazil-ui-overview} A) controls the animation process.
CoronaViz can be run in two animation modes:  ``Total'' and ``Window''.
In Window Mode we provide a temporal region (termed the ``Animation Window'') that is specified in terms of a range of days. Window Mode further filters the data within the spatial range (i.e. region) being viewed, creating a spatiotemporal query.
To simplify the explanation, we use a variant of the example query of animating the progression of COVID-19 in Brazil (See Fig.~\ref{fig:brazil-ui-overview} for its result) and its neighboring South American countries in terms of the values of disease-related variables.
the year 2020 though October 12 of the same year.
The animation can provide either the cumulative values of these variables or the daily average value for the days making up the window.
Note that when the window duration is one day, then the cumulative value and the average daily rate are the same.
This information is provided on a daily basis on the last day of the animation window for each day of the animation range.
In contrast, recall that the maximum possible size of the animation window is the duration of the entire timeline.
This case yields no actual effective animation as the result is the cumulative value of the variables and the daily average value of the variables over the entire ``Animation Range'' and only reported on the last day of the animation window.
Users don't have to know the value of this maximum as it is specified by checking the ``Maximum'' checkbox that appears to the right of the ``Window Size'' in the animation panel. 

Users who wish to see the cumulative as well as the average daily rates for all of the disease-related variables in an animated manner can use the ``Total'' mode.
In this case, we do have an animation on a daily basis with the final frame of the animation yielding the cumulative values of the disease-related variables for the temporal ``Animation Range'' for all spatial ranges that can be viewed.
These features are all accessed by clicking on the ``Animation'' tab in the ``Control Panel.'' 
Figure~\ref{fig:brazil-ui-overview} is the final screenshot for the animation of the cumulative values of the number of confirmed cases, deaths and recoveries for the Total mode query for the countries in the vicinity of Brazil for the selected time period.
Note the larger geocircles on account of no normalization which is the case when we used rates for the disease-related variables.

The ``Location'' tab (Fig.~\ref{fig:brazil-ui-overview} B) activates the ``Location Specification'' process which identifies the spatial entity for which we wish to animate and view the disease-related variables.
This location is known as the ``Animation Focus.''
It can be the name of a country/region, state/province, or county/city all of which are
obtained from an appropriately named pull-down menu.
Alternatively, the geographical entities can also be specified graphically using direct manipulation actions like pan, zoom, and hover.
In this case we usually start with a map from which a new map is constructed using pan and zoom operations as well as possibly dilation.
Once the desired location has been identified on the map, then a single left click on the mouse is sufficient to initialize or reset the ``Animation Focus.''

The advantage of the direct manipulation approach is that it provides the query poser the opportunity to specify the exact shape and boundary, as well as the resolution, of the query region.
The ``Location'' tab can also be used in the same manner to set what we call a ``Baseline Location'' for comparing disease-related data as the animation proceeds. This location can only be set using the pull-down menus and cannot be set using direct manipulation.
As the animation proceeds, the values of all of the disease-related variables and rates are displayed side-by-side in the ``Animation Focus Data Panel'' for the two locations.

At this point, the animation can be started by clicking on the ``Start Animation'' button in the ``Animation Control Panel.''
Brazil is said to be the ``Focus Location'' which means that as the animation proceeds, users can see additional data for Brazil corresponding to the daily variation of of all of the disease-related variables and rates by looking at the panel with the heading ``Animation Focus Data.'' (Fig.~\ref{fig:brazil-ui-overview} E).

\begin{figure*}[ht!]
    \centering
    \subfloat[European countries (Mar.-Apr.)]{
        \includegraphics[width=0.45\textwidth,trim=0cm 3cm 0cm 2cm,clip]{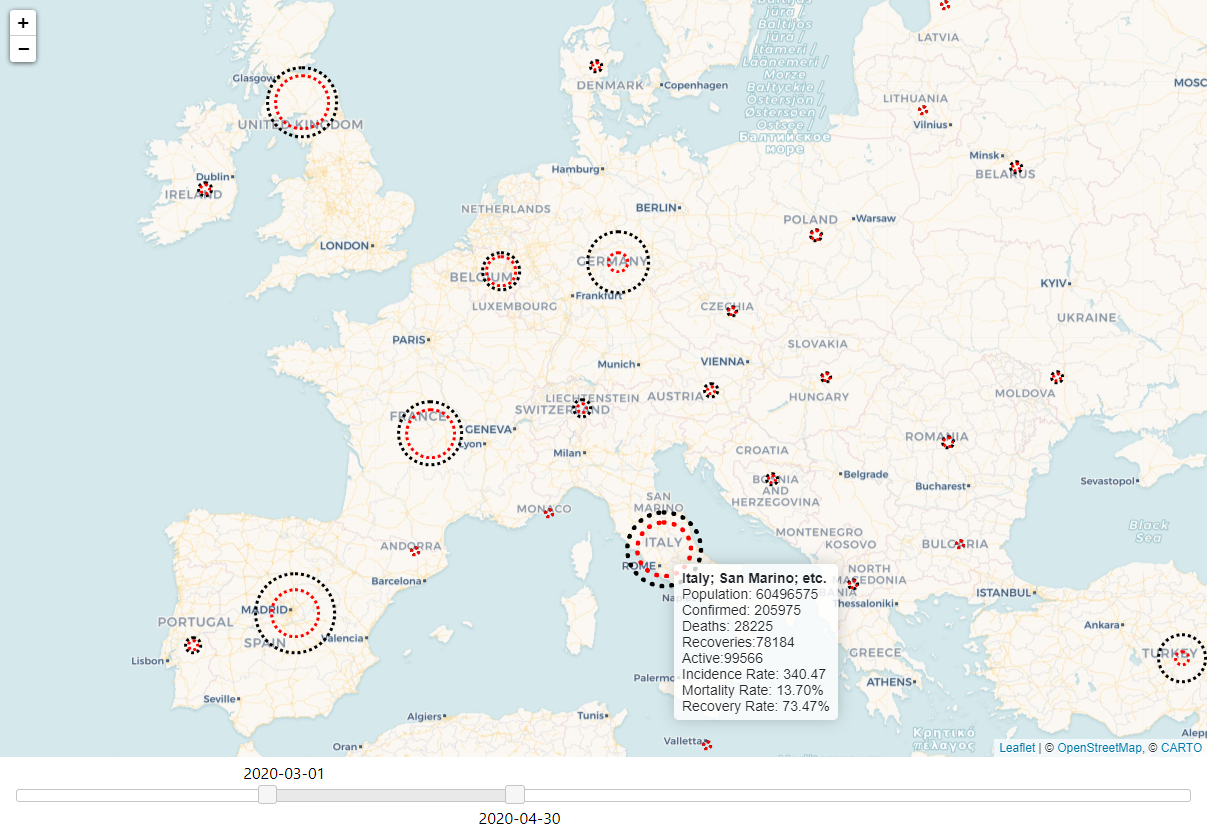}
        \label{fig:europe}
    }
    \subfloat[Northern Europe (Jan.-Oct.)]{
        \includegraphics[width=0.45\textwidth,trim=0cm 3cm 0cm 2cm,clip]{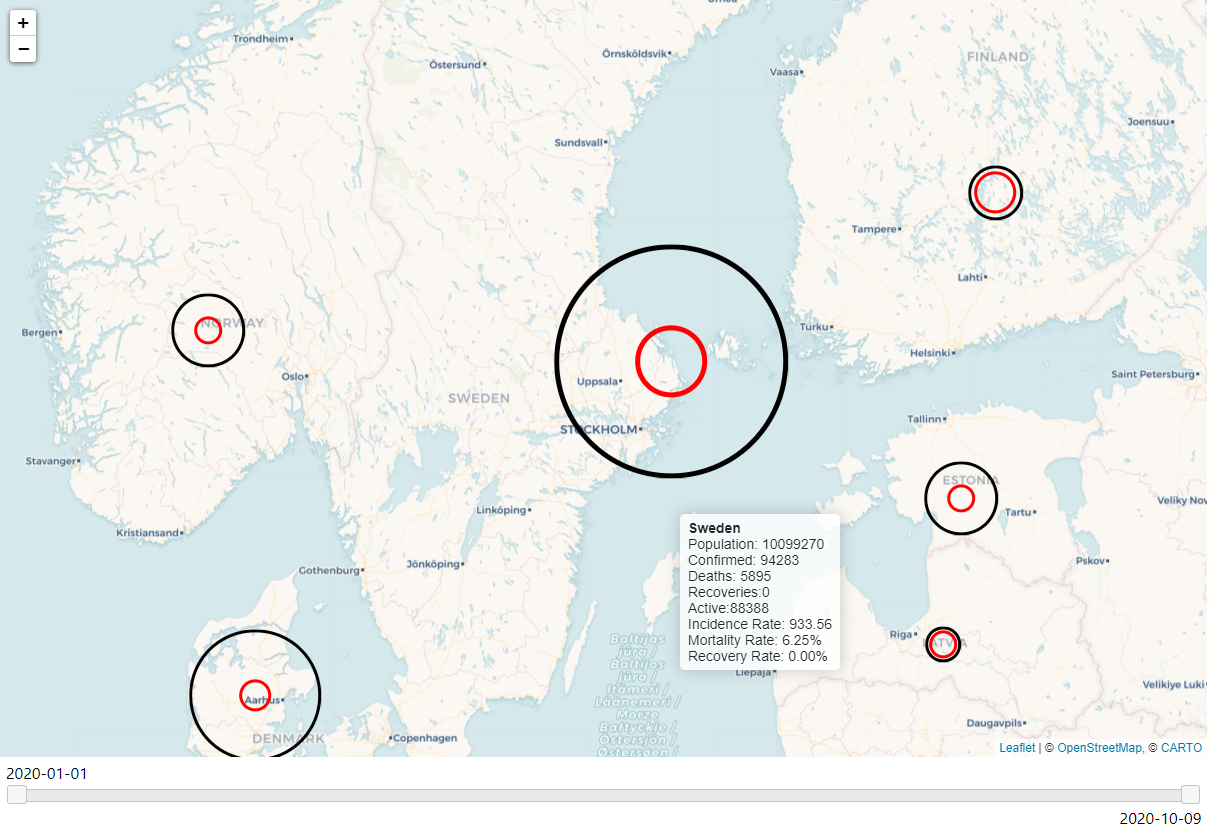}
        \label{fig:sweden}
    }\\
    \subfloat[Counties near Washington D.C. (Jan.-Oct.)]{
        \includegraphics[width=0.45\textwidth,trim=0cm 3cm 0cm 2cm,clip]{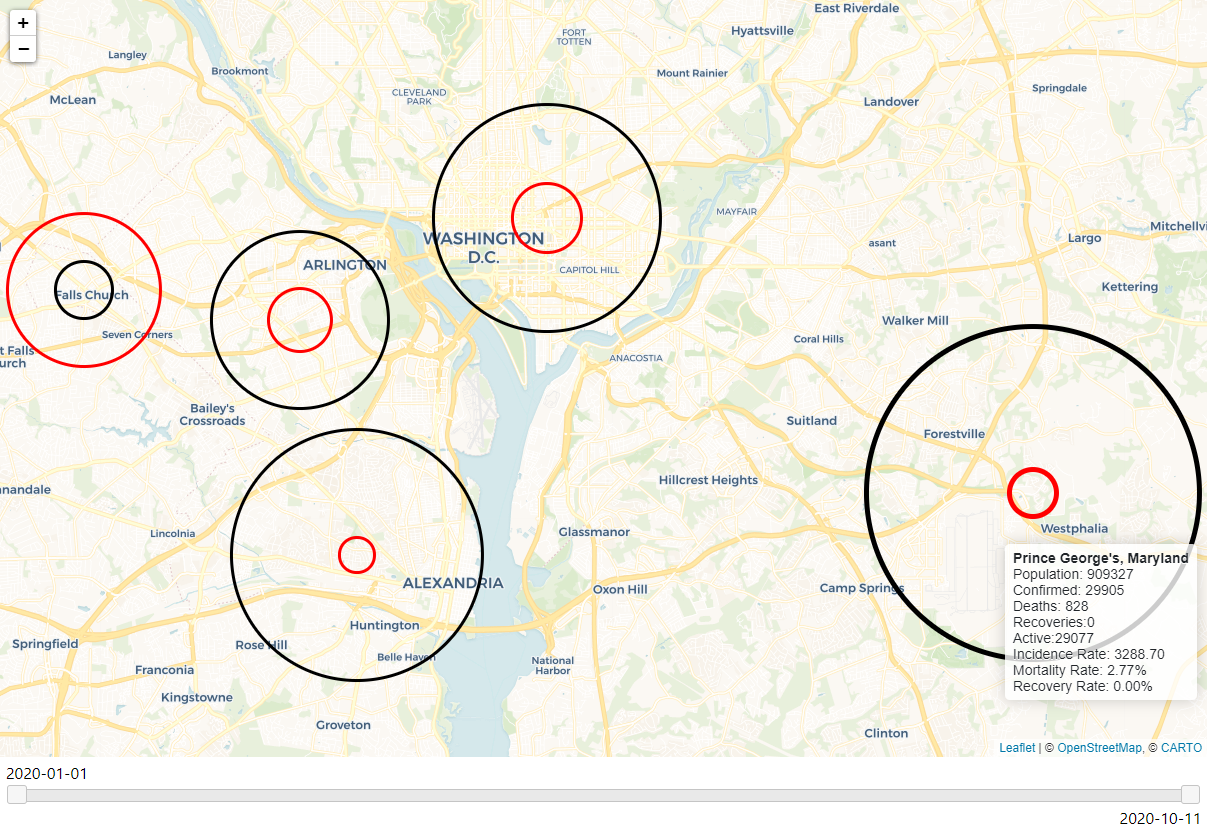}
        \label{fig:counties}
    }
    \subfloat[Sweden and Israel.]{
        \includegraphics[width=0.45\textwidth,trim=0cm 0cm 0cm 0cm]{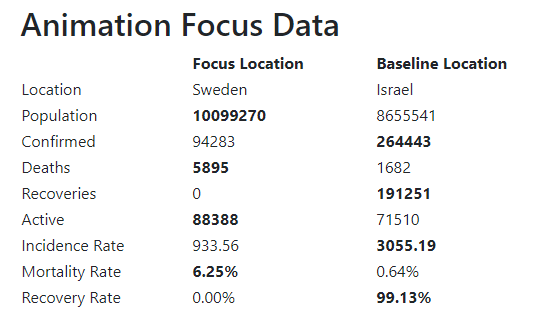}
        \label{fig:sweden-israel}
    }
    \caption{Case studies~(\href{https://www.youtube.com/watch?v=QSkI8htZQQo}{\color{blue}\underline{video}})}
    \label{fig:case}
\end{figure*}

\subsection{View Management and Filtering}

The ``View'' tab (Fig.~\ref{fig:brazil-ui-overview} C) in the ``Control Panel'' controls the viewing process by providing a number of options of viewing the different disease-related rates and variables.
The user selects them by clicking the checkbox to their right.
Some combinations of variables are pre-defined such as the ``Default'' view corresponding to just displaying the incidence and mortality rates, while the ``Rate'' view corresponds to just displaying the incidence, mortality, and recovery rates.
As the animation proceeds, the values of the selected disease-related variables and rates are displayed using geocircles anchored at the corresponding geographic location and whose radii provides an indication of their relative magnitudes.

\subsection{Geocircle Scaling Options}
We have several scale options for the radii: \textit{linear}, \textit{logarithmic}, and \textit{Flannery}, all of which have different benefits.
Using a \textit{linear} relationship makes the radii directly proportional to values. Note that this would cause the area of the circles to have a squared relationship with values, which could be misleading. This can also make it difficult to simultaneously view values with wide ranges (say, differing by more than one order of magnitude). However, we provide the option as it can be useful for discerning subtle differences among regions for which a value is in a similar range.
A \textit{logarithmic} relationship makes it possible to view much wider ranges in values, but at the expense of seeing finer differences.
Another natural option would be to make the area, rather than the radius, of the geocircles proportional to variable values. This would be equivalent to scaling radii by an exponent of 0.5 (i.e. the square root). However, the visual system is not good at judging the relative values of areas~\cite{cleveland1984graphical}. Instead, we offer the option of \textit{Flannery} scaling, which uses a psychophysically determined exponent of .57 to scale each dimension of a proportional symbol~\cite{flannery1971relative}. This has the disadvantage that there is no intuitive relationship for the user between the rendered sizes and the underlying data. However, as Flannery scaling is widely used in cartography, we make this option the default, while offering others for advanced users.
On top of the above translations, an additional linear scaling factor from .1X to 8.0X can be applied to each variable using sliders (Fig.~\ref{fig:brazil-ui-overview} C).

\subsection{Spatial Queries}

In addition, during the animation, the mouse can be moved over the visible part of the map (termed ``hover'') and the data associated with closest geocircle (e.g., Peru in Fig.~\ref{fig:brazil-ui-overview} F) is  displayed in what we call the ``Hover Box,'' which is anchored on mouse and moves with the pointer. This nearest geocircle is highlighted with a thicker outline.

Figure~\ref{fig:brazil-ui-overview} shows the result of the animation for a time period, which is set in the ``Animation Control'' panel.
Note that the ``Animation Control'' panel enables users to pause, resume, halt, and restart the animation process by clicking on the appropriate button.
Users can also run the animation in a day-by-day manner one day at a time in the forward and backward temporal directions via the buttons labeled ``$<$'' and ``$>$'', respectively.
It is especially interesting to go backwards at the end of the animation by repeatedly clicking on the ``$<$'' button found to the left of the ``Start/Pause/Resume/Stop'' button.
The above playback can be achieved in a continuous manner by using the mouse to define the width of a window by varying the positions of the left and right boxes of the time slider (Fig.~\ref{fig:brazil-ui-overview} G). This process proceeds by fixing the right box and varying the left box as needed. The playback is achieved by dragging the left box in either of the two temporal directions. During the animation, the right box follows the motion of the the left box.

\subsection{Implementation}

The map query interface is built around an interactive web map provided by the Leaflet JavaScript Library~\cite{leaflet} and the OpenStreetMap API~\cite{OpenStreetMap} for geospatial rendering.
Data are periodically retrieved from Johns Hopkins University~\cite{govex,csse} in CSV format are parsed and data values are aggregated by location using Pandas~\cite{reback2020pandas} and Numpy~\cite{harris2020array}.
Marker clustering~\cite{markerCluster} is implemented using an extension to Leaflet.  

\section{Examples}
\label{util}

In this section we provide use cases of CoronaViz that demonstrate its utility.
Notice that we provide both figures and animations.
The figures usually correspond to the last frame of an animation.
In most cases we also provide a link to  video for the entire animation.

Spatial data values sometimes have very small magnitude.
This is the case for mortality rates, especially when compared with the number of confirmed cases or incidence rates, thus making it difficult to compare their values for different locations.
Users can perform more meaningful comparisons of locations with very small data values by changing the scaling factor.\footnote{\url{https://www.youtube.com/watch?v=VLiWoWtYHQo}}
For example, we see the mortality and incident rate for the United States for September 2020 (window mode).
Since the variables are relatively small, consequently, the geocircles representing them are also small.
By changing the scaling factor of the circles, their sizes become larger simultaneously.
After that, the differences in the mortality and incident rate between two locations are clearer.

Besides using raw data directly, we often also normalize the data based on population.\footnote{\url{https://www.youtube.com/watch?v=cCGWQ4jaChw}}
Some countries like the U.S.\ and Brazil have a large population and thus they have many
confirmed cases, which results in geocircles with large radii when the raw data is plotted directly.
After normalization, the values of the confirmed cases are represented by the incidence rate, which is defined as the number of confirmed cases per 100,000 population.
The incidence rate is rarely greater than 3,000 and hence the values of the radii of the geocircles become reasonable after normalization.

Users can control the number of geocircles on the map.\footnote{\url{https://www.youtube.com/watch?v=DYHk5XmGXKA}}
Geocircles aggregate automatically if they are close to each other, by using marker clustering, which allows large numbers of points to be rendered quickly without overloading the user with information.
As the user increases the zoom level of the map, focusing on a smaller area, these aggregate markers are split into two or more new markers that together represent all the points represented by the original marker. This decomposition allows more detail to be displayed when examining a small area without showing too much detail at lower zoom levels. For example, in Figure~\ref{fig:dc-maryland-aggregate} the geocircles of Washington, D.C. and Maryland usually aggregate unless we zoom in substantially, since they are geographically close.
By increasing the number of geocircles plotted on the map, close geocircles (e.g., D.C. and Maryland in Fig.~\ref{fig:dc-maryland-separate}), and more details are shown on the map.
This makes comparisons among geographically-proximate locations feasible.

We also study some typical cases to show the utility of CoronaViz. In Europe, the pandemic first peaked in late March to early April. As shown in Figure~\ref{fig:europe} (confirmed cases and deaths), there were several hot spots. Setting the temporal window to be March and April finds them to be the U.K., France, Germany, Italy, and Spain. Another example is Sweden, which let the Coronavirus spread in the hope that the population would develop ``herd immunity.'' Figure~\ref{fig:sweden} shows the incidence and mortality rates for Sweden and its neighboring countries for January through October 2020 (Total mode).
From Figure~\ref{fig:sweden}, we see that Sweden has higher incidence and mortality rates than its neighboring countries. We can also compare the data through the text information provided in the sidebar.\footnote{\url{https://www.youtube.com/watch?v=QSkI8htZQQo}}
In Figure~\ref{fig:sweden-israel}, we use Israel as a baseline, whose population is close to Sweden.
We observe that Israel has a higher incidence rate but a lower mortality rate compared with Sweden.
Note that we do not have recovery data for Sweden, so it is not shown in Figure~\ref{fig:sweden-israel}. Observe that CoronaViz not only visualizes data of countries but also other administrative divisions
like states and counties.
For example, some counties near Washington D.C.\ are plotted in Figure~\ref{fig:counties} for January through October 2020 (Total mode).

\section{User Study} \label{us}

To understand the utility of animated geocircles and the CoronaViz system, we conducted a user study in which we compared CoronaViz to two other coronavirus tracking systems: the \textit{New York Times}’ dashboard~\cite{tool-nyt} (hereafter referred to as NYT) and the Johns Hopkins University’s dashboard~\cite{tool-jhu} (hereafter referred to as JHU).
Our goal was to assess the value of a combined spatiotemporal visualization, as opposed to a dashboard approach with geospatial and temporal information captured by discrete visualizations.

\subsection{Participants}

Twelve participants were recruited for this study. The participants were gathered from the University of Maryland’s computer science graduate students’ emailing list.
We had two participants whose age was between 18--25, 9 participants between 26--35, and one participant between 51--70.
Four participants completed or are completing a masters or professional degree, the other 8 completed or are completing a terminal degree (PhD/MD/JD).
While recruiting, we asked for all participants willing and able to conduct an hour-long Zoom session.
The users needed to have a stable internet connection, the Safari or Chrome browser, and the Zoom application.

\subsection{Tasks and Hypotheses}
\label{sec:task-hyp}
During this study we asked questions requiring users to perform tasks that fall under 3 types of queries, all centering around U.S.\ states:

\begin{itemize}
    \setlength{\itemsep}{0cm}
    \item \textbf{Task 1}: Spatial proximity, e.g. ``Find the highest cumulative case count for a state that borders [\textbf{state}] using [\textbf{tool}].''
    \item \textbf{Task 2}: Temporal proximity, e.g. "Determine whether the 7-day average of cases is generally increasing or decreasing for [\textbf{state}] during the month of [\textbf{month}] 2020 using [\textbf{tool}].''
    \item \textbf{Task 3}: Spatiotemporal proximity, e.g. ``Beginning from [\textbf{date}], using the 7-day confirmed case count, when is [\textbf{state}] first surpassed by a state that it shares a border (or corner) with, and which state(s) was it? Use [\textbf{tool}].''
\end{itemize}

\noindent In formulating hypotheses, we consider that multidimensional visualizations (e.g. those that include both time and space) must used varied encodings, and thus must sacrifice some of the efficacy of those encodings for some dimension or dimensions. Thus, while we posit that a unified visualization would have utility for queries across time and space, a query of a single dimension would be better served by a dedicated visualization of the type employed by dashboards.
Accordingly, we formulated 3 hypotheses, corresponding to the three tasks:

\begin{itemize}
    \setlength{\itemsep}{0cm}
    \item[\textit{H1}] Users will deem CoronaViz less useful than JHU and NYT for spatial proximity queries (Task 1). 
    \item[\textit{H2}] Users will deem CoronaViz less useful than JHU and NYT for temporal proximity queries (Task 2).
    \item[\textit{H3}] Users will deem CoronaViz more useful than JHU and NYT for spatiotemporal proximity queries (Task 3).
\end{itemize}

\subsection{Experimental Design}

We used a within-subjects design, with each participant trying all three tasks with all three tools.
This meant that within each task, the query scenario (time/location) had to be varied so participants could not carry over answers found with one tool to another tool.
We thus developed three scenarios of similar difficulty for each task.
So that tools were not penalized by variance in scenario difficulty, the scenarios were rotated to form three versions of the complete questionnaire, each with different tool-scenario combinations.

Since all three tools are complex platforms with many ways to perform tasks, measurements of time and accuracy of tasks could be confounded by factors such as exploration of the tool, their understanding of the interface, and system responsiveness.
Modifying the tools or tasks to constrain them, however, would likely result in contrived scenarios that would not be ecologically valid.
As a way around this problem, we chose to measure the users’ own assessments of how ``useful'' each tool would be under proposed scenarios.
These scenarios were constructed to correspond to the three types of tasks that the users had performed during the study, allowing them to extrapolate.
Our main dependent variable is thus Likert-scale responses from the post-study survey, which asks the user to estimate the usefulness of each tool for each hypothetical task.
We also ask, for each task type, which of the three tools they would choose to perform the task.
This provides a secondary dependent variable.
Following the choice of preferred tool for each task, users were asked to explain in free-form text why they would choose that tool.
Finally, we asked several more free-form text questions about what was good and not good about each tool, and any other comments.

\subsection{Procedure}

Each of the participants were given one hour to complete the study.
For consistency, at the start of the study, we allocated 5 minutes for a video demonstration of each tool. 
Thereafter, the users were asked to share their screen and start answering the questionnaire.
The first two questions were \textit{training questions}, which were given to help participants to become familiar with each tool and were not scored.
One training question focused on a basic geospatial query, while the other focused on a basic temporal query. Participants completed both training questions for all three tools. For NYT and JHU, this ensured that participants were familiar with the separate visualizations of space and time before starting the real tasks. For CoronaViz, it ensured that they grasped how to make both spatial and temporal queries within the unified interface, using the map view and the animation controls, respectively.
While answering the training questions, participants had the opportunity to ask questions about the platform and how to do the task, and researchers running the study offered guidance if a participant was struggling.
After training, users were asked to answer the \textit{study questions}.
During these, the above interactions were not permitted except for simple task clarifications.
There were nine study questions, corresponding to the use of all three tools to perform each of the three tasks outlined in Section~\ref{sec:task-hyp}.
Towards the end of the study, users were given a 5-minute warning.
As time expired, they were asked to complete the post-study survey on their own time, which thereafter, they were compensated via a \$15 Amazon gift card.
The compensation is based on the minimum wage for an employee in the state of Maryland and on the basis of Amazon’s accessibility to all.

\section{Results}

\begin{figure}[t]
    \centering
    \includegraphics[width=0.75\columnwidth]{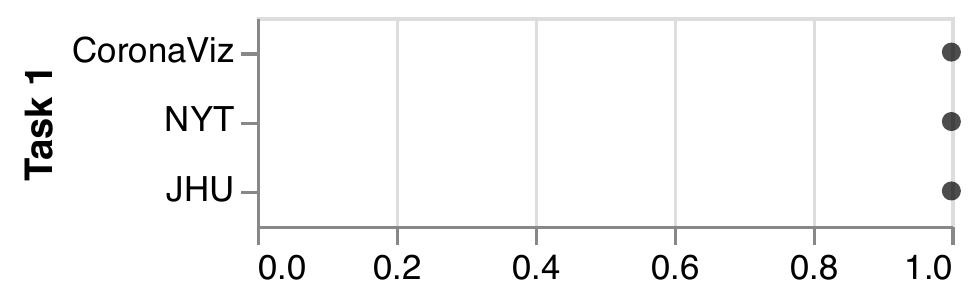}
    \includegraphics[width=0.75\columnwidth]{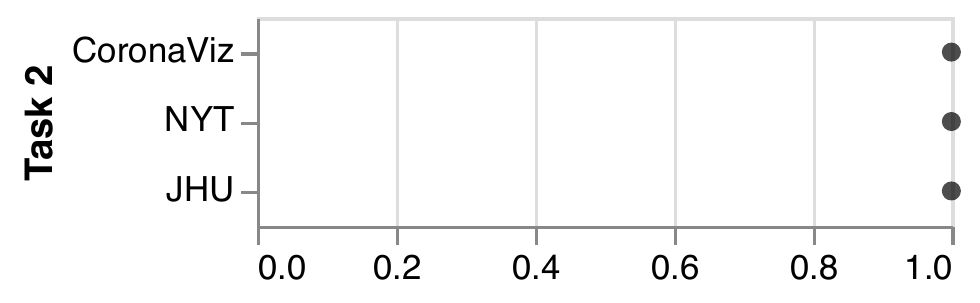}
    \includegraphics[width=0.75\columnwidth]{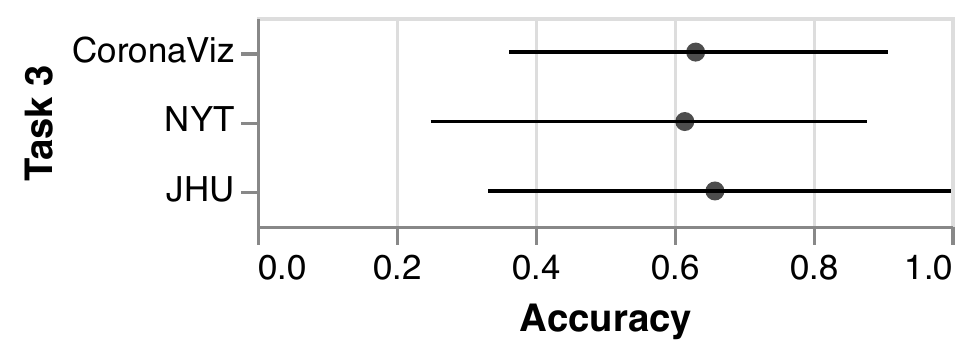}
    \caption{Accuracy by task and tool.
    Dots represent means and horizontal lines represent 95\% confidence intervals.}
    \label{fig:accuracy}
\end{figure}

\begin{figure}[t!]
    \centering
    \includegraphics[width=0.75\columnwidth]{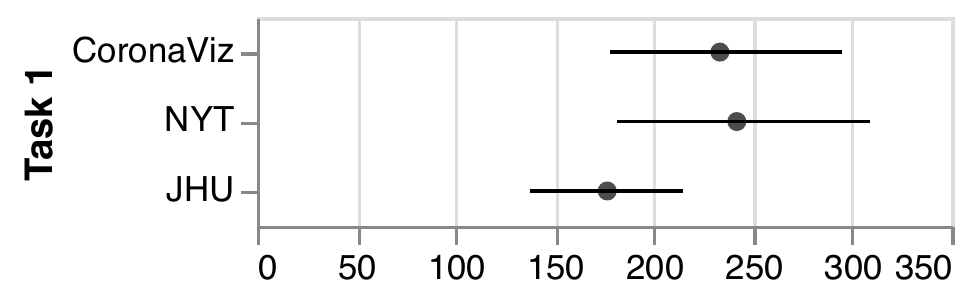}
    \includegraphics[width=0.75\columnwidth]{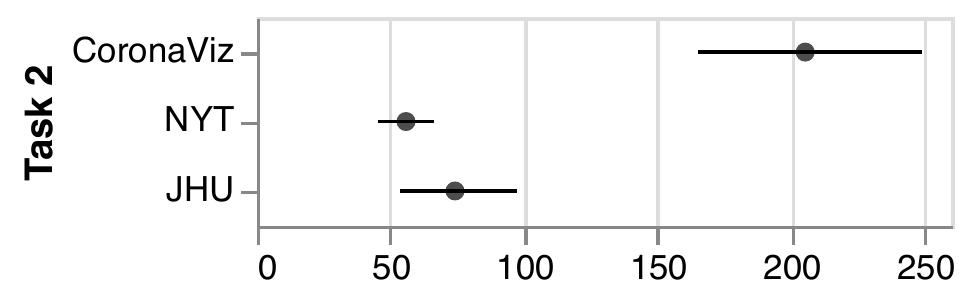}
    \includegraphics[width=0.75\columnwidth]{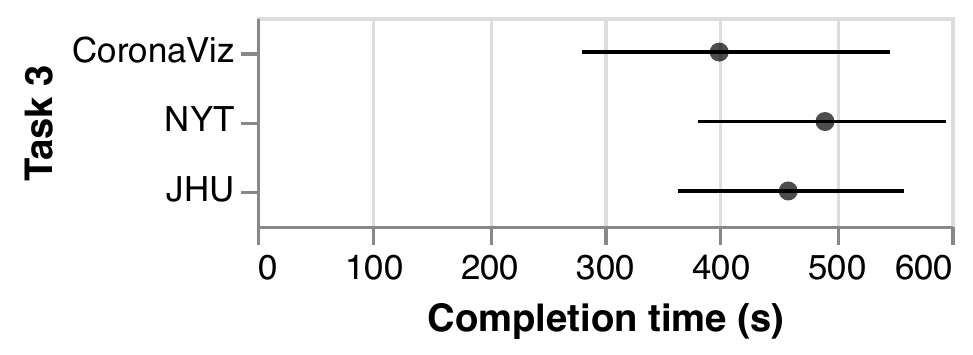}
    \caption{Task completion time by task and tool.
    Dots represent means and horizontal lines represent 95\% confidence intervals.}
    \label{fig:time}
\end{figure}

\begin{figure*}[ht!]
    \centering
    \includegraphics[width=\textwidth]{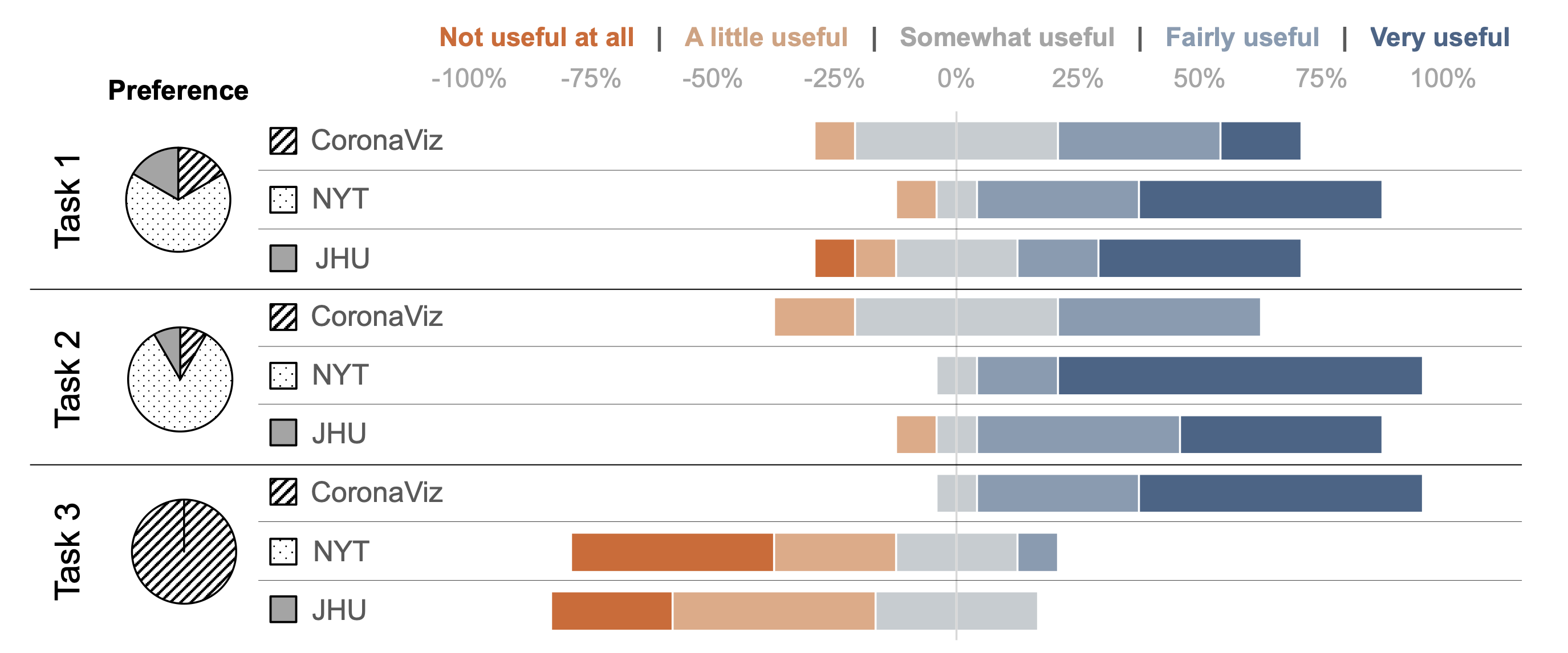}
    \caption{User study survey results by task. Left, responses to the question of which tool the participants would choose if asked to perform a task similar to the one indicated.
    Right, proportions of Likert-scale responses to the question of how useful each tool would be for performing a similar task, shown as stacked bars centered around neutral (``Somewhat useful'') responses.}
\label{fig:likert}
\end{figure*}

Though user judgments of tool usefulness were our main variable, we also assessed how fast and accurate users were for each combination of task and tool and analyzed text responses.

\subsection{User Assessment of Utility}

Figure~\ref{fig:likert} shows a summary of Likert-scale responses for how useful each tool was judged to be for hypothetical tasks representing spatial, temporal, and spatiotemporal proximity queries (right) and which tool users would choose to perform each query (left).

\subsection{Time and Accuracy}

Time was manually coded by reviewing recordings of the sessions and marking when users started and completed each task.
Accuracy was computed as a binary (0 or 1) for whether the response the user entered in the questionnaire was correct.
Missing values for incomplete tasks were excluded.
Figure~\ref{fig:time} shows mean times for each combination of task and tool, with 95\% confidence intervals, computed via bootstrapping.
Figure~\ref{fig:accuracy} shows mean accuracy for each combination of task and tool, with 95\% confidence intervals computed via bootstrapping.

\section{Discussion}

The main variables we tested (user judgment of tool usefulness and preferred tool) supported our hypotheses for all three tasks.
Recurring themes in the free-form text responses from participants also support our initial reasoning for these hypotheses. 

\subsection{Hypothesis 1: Spatial Proximity Queries}

For H1 (that users would find CoronaViz less useful than the other tools for a purely spatial proximity query), Likert-scale responses of assessed utility are higher for JHU and NYT than for CoronaViz.
This is not surprising, as both JHU and NYT have more straightforward routes to access data for this query. 
Both tools contain, among the other items in their dashboards, tables of values for each state.
This made the task a straightforward lookup of values once the list of neighboring states was acquired from the map.
NYT was widely chosen as the preferred tool for this task, likely because of this direct access to data.
This is supported by comments such as:

\begin{itemize}
    \setlength{\itemsep}{0cm}
    \item (choosing NYT) ``...compared to CoronaViz, [NYT] shows the numbers directly. When values are close to each other [in CoronaViz], hard to tell the difference based on sizes of circles, then you need hover on each circle to see the number.''
    \item (choosing NYT) ``[NYT] provides a sorted list of [states] straightaway.''
    \item (choosing NYT) ``[NYT] had a list of items that reported the values and thus enabling a faster answer.''
    \item (choosing JHU) ``The admin panels [of JHU] let you drill down on a state quickly, and counties are sorted automatically by case''
\end{itemize}

\subsection{Hypothesis 2: Temporal Proximity Queries}

For H2 (that users would find CoronaViz less useful than the other tools for a purely temporal proximity query), Likert-scale responses of assessed utility are much higher for JHU and NYT than for CoronaViz.
This is also not surprising, as both JHU and NYT have dedicated time series plots.
While time-series plots cannot directly integrate spatial information the way CoronaViz does using animation for time, these plots make pure temporal queries much more straightforward.
This is supported by comments such as:

\begin{itemize}
    \setlength{\itemsep}{0cm}
    \item (choosing NYT) ``[NYT] had a plot, I can look for a spike easily in that plot.''
    \item (choosing NYT) ``once you get the state, the confirmed cases over 7 days chart is brought to the center and the top is marked, so super easy to spot.''
    \item (choosing NYT) ``[NYT] has a time series curve that can help find the maximum/minimum very easily''
    \item (choosing JHU) ``Graph is discrete and can easily pick out which day without having to hover each day or running any animations''
\end{itemize}

Ten of twelve users said they would prefer to use NYT for this query, with one preferring JHU and one preferring CoronaViz. 
Since both NYT and JHU have similar time series plots, we conjecture that the large skew in preference toward NYT comes from differences in the interface.
This is supported by comments such as:

\begin{itemize}
    \setlength{\itemsep}{0cm}
    \item (choosing NYT) ``Both NY Times and JHU provide a time plot. But JHU one is kinda small and need to take an extra action to zoom in.''
    \item (choosing NYT) ``It's very easy to find a state and it doesn't slow down my browser like JHU does''
\end{itemize}

\subsection{Hypothesis 3: Spatiotemporal Proximity Queries}

Finally, for H3 (that users would find CoronaViz more useful than the other tools for a spatiotemporal proximity query), Likert-scale responses of assessed utility are heavily skewed toward CoronaViz.
All twelve participants responded that they would choose CoronaViz over the other tools for this type of task.
We submit that this preference comes from the many, cumbersome steps that are required to make comparisons in both spatial and temporal dimensions simultaneously without the use of animation.
This is supported by comments such as:

\begin{itemize}
    \item (choosing CoronaViz) ``looking at bordering states are easier to locate in [CoronaViz], [JHU] gives neighboring state but needs back and forth scrolling and [NYT] needs a separate map and comparison is hard as there is not side by side info''
    \item (choosing CoronaViz) ``Use animation to narrow down which state and around which day, then pick which one''
    \item (choosing CoronaViz) ``Both [NYT] and [JHU] dont show the data of multiple locations at the same time. You need open multiple windows to answer this question.''
    \item (choosing CoronaViz) ``[CoronaViz] allows for comparison between the states with animation controls boldening the relatively higher number of cases.''
\end{itemize}

\subsection{Time and Accuracy}

Both time and accuracy were similar for all tools for each task, with the exception of the time taken to complete Task 2 with CoronaViz, which is strikingly higher than the other tools.
This is not surprising, as encoding the temporal dimension using sequential renderings has the drawback that the entire timeline cannot be seen at one time, adding steps to comparison.
However, the value of this encoding is clearer when the interaction of the temporal and spatial dimensions of variables needs to be investigated, as in Task 3. 

\subsection{Study Limitations}

While our study showed several interesting results, it is not definitive and had aspects that were not ideal.
First, for consistency and fairness to participants we capped total user study time at one hour.
This resulted in some participants not finishing all queries with all three tools.
However, all participants were able to complete Task 1 (spatial proximity query) and Task 2 (temporal proximity query) with all three tools.
Eleven of the twelve participants completed Task 3 (spatiotemporal proximity query) using at least CoronaViz, and nine of the twelve at least attempted Task 3 with CoronaViz and one other system.
Incomplete results were excluded from analysis of time and accuracy.
However, we deemed all participants able to extrapolate from their experience with the tools how easy or hard it would be to perform the hypothetical tasks with them, even if they did not complete one of the tasks.
We thus included survey results from all participants.

Second, due to the in-depth, one-on-one nature of our study, we had a relatively small pool of participants and thus did not achieve complete combinatorial ordering of conditions.
We chose scenario (i.e., the location and time being queried) as the most important factor to rotate, since variations in border topology and case dynamics are likely to effect difficulty.
This meant participants tried each of the three tools in the same order for each task.
We addressed this by ordering CoronaViz before the other tools for each task, preventing  learning effects from influencing the perceived strength of our own tool.


\section{Conclusion and Future Work}
\label{c}

In developing CoronaViz, we sought to provide a more complete sense of the spread of the COVID-19 pandemic by creating a unified spatiotemporal display with support for viewing multiple variables simultaneously. In contrast to existing tools, this allows for making queries simultaneously across both temporal and geospatial ranges. The ability to query and animate temporal ranges, rather than single points in time, is an important interactive element when viewing COVID-19 data.\footnote{Personal communication with Terry Slocum, University of Kansas, 2021.} We have shown that this unified system can be of value to users when making queries with both spatial and temporal components. Our visualization relies heavily on the availability of quantitative data about the presence of the disease provided by the Johns Hopkins University. In the future, CoronaViz could easily incorporate additional variables as they become available, such as hospitalization rates, which are currently not available due to concerns over the reliability and stability of the reported data~\cite{glassman2022hospitalization}. A similar problem exists with test positivity data; the number of tests is unevenly reported thereby making it impossible to report this rate accurately. We plan to continue to update CoronaViz to include these data as they are available.

Additional useful knowledge about the potential progression of the disease can be gained by keeping track of
spatially-referenced mentions in news articles as in NewsStand~\cite{Lan14,Lieb12b,Same14}, tweets as in TwitterStand~\cite{Gram13,Jack11,Sank09c}, documents such as PubMed~\cite{Lieb07} and ProMED-mail~\cite{Lan12,Lieb07}, and spreadsheets~\cite{Adel13}.
This involves geotagging which is the process of recognizing textual references to location~\cite{Quer10,Lieb10b}.
Presently we do not make use of such data although we do feel that such an approach is a direction for future research. 

Finally, while the features in CoronaViz were developed specifically to better understand the COVID-19 pandemic, we note that the use of animated geocircles could have just as much value for tracking any epidemic or pandemic, such as seasonal influenza. It is our hope that making our tool open-source and publicly hosted will spur further development in the field of interactive visualization for epidemiology.

\section*{Acknowledgments}

This work was supported partly by grants IIS-18-16889, IIS-20-41415, and IIS-21-14451 from the U.S.\ National Science Foundation.
Any opinions, findings, and conclusions expressed in this material are those of the authors and do not necessarily reflect the views of the funding agency.

\bibliographystyle{IEEEtran}
\bibliography{hjs,corona_tracking,tools}

\begin{IEEEbiography}[\framebox{\resizebox{1in}{1.25in}{?}}]
{Brian Ondov}
received the Ph.D. degree in Computer Science in 2021 from the University of Maryland, College Park in College Park, MD, USA.
He is a postdoctoral researcher at the National Library of Medicine at the National Institutes of Health in Bethesda, MD, USA.
\end{IEEEbiography}

\begin{IEEEbiography}[\framebox{\resizebox{1in}{1.25in}{?}}]{Harsh B. Patel}
received the B.S.\ degree in computer science in 2019 from Purdue University in West Lafayette, IN, USA.
He is a masters student in computer science at University of Maryland, College Park in College Park, MD, USA.
\end{IEEEbiography}

\begin{IEEEbiography}[\framebox{\resizebox{1in}{1.25in}{?}}]{Ai-Te Kuo}
received the M.S.\ degree in computer science in 2019 from Auburn University in Auburn, AL, USA.
He is a Ph.D. student in computer science at Auburn University.
\end{IEEEbiography}

\begin{IEEEbiography}[\framebox{\resizebox{1in}{1.25in}{?}}]{Hanan Samet}
received the Ph.D.\ degree in computer science in 1975 from Stanford University in Stanford, CA, USA.
He is a distinguished professor in the Department of Computer Science at University of Maryland, College Park in College Park, MD, USA.
He is a Life Fellow of the IEEE.
\end{IEEEbiography}

\begin{IEEEbiography}[\framebox{\resizebox{1in}{1.25in}{?}}]{John Kastner}
received the M.S.\ degree in computer science in 2021 from University of Maryland, College Park in College Park, MD, USA.
He is an applied scientist at Amazon, Inc.
\end{IEEEbiography}

\begin{IEEEbiography}[\framebox{\resizebox{1in}{1.25in}{?}}]{Yungheng Han}
is a Ph.D.\ candidate in computer science at the University of Maryland, College Park in College Park, MD, USA.
\end{IEEEbiography}

\begin{IEEEbiography}[\framebox{\resizebox{1in}{1.25in}{?}}]{Hong Wei}
received the Ph.D.\ degree in computer science in 2020 from University of Maryland, College Park in College Park, MD, USA.
He is a research scientist at Facebook, Inc.
\end{IEEEbiography}

\begin{IEEEbiography}[\framebox{\resizebox{1in}{1.25in}{?}}]{Niklas Elmqvist}
received the Ph.D.\ degree in 2006 from Chalmers University of Technology in G\"{o}teborg, Sweden.
He is a professor in the College of Information Studies, University of Maryland, College Park in College Park, MD, USA. 
He is also a member of the Institute for Advanced Computer Studies (UMIACS) and formerly the director of the Human-Computer Interaction Laboratory (HCIL) at UMD.
He is a senior member of the IEEE and the IEEE Computer Society.
\end{IEEEbiography}

\vfill

\end{document}